\documentclass[aps,prl,reprint,groupedaddress]{revtex4-1}

\usepackage{graphicx}  
\usepackage{booktabs}
\usepackage{multirow}
\usepackage{color}

\usepackage{dcolumn}   
\usepackage{bm}        
\usepackage{amssymb}   
\usepackage{amsfonts}
\usepackage{verbatim}
\usepackage{amsmath}
\usepackage{amsfonts}
\usepackage{ulem}
\newcommand{\rfm}[1]{\color[rgb]{0,0,0}#1}

\def\lsim{\mathrel{\rlap{\lower4pt\hbox{\hskip1pt$\sim$}}
    \raise1pt\hbox{$<$}}}                
\def\gsim{\mathrel{\rlap{\lower4pt\hbox{\hskip1pt$\sim$}}
    \raise1pt\hbox{$>$}}}                

\begin{document}
\normalem

\title{Phonon-Mediated Quasiparticle Poisoning of Superconducting Microwave Resonators}

\author{U. Patel$^{1}$}
\author{Ivan V. Pechenezhskiy$^1$}
\author{B. L. T. Plourde$^{2}$}
\author{M. G. Vavilov$^{1}$}
\author{R. McDermott$^{1}$}

\email[Electronic address: ]{rfmcdermott@wisc.edu}

\affiliation{$^{1}$Department of Physics, University of Wisconsin, Madison, Wisconsin 53706, USA}
\affiliation{$^{2}$Department of Physics, Syracuse University, Syracuse, New York 13244, USA}

\date{\today}

\begin{abstract}
Nonequilibrium quasiparticles represent a significant source of decoherence in superconducting quantum circuits. Here we investigate the mechanism of quasiparticle poisoning in devices subjected to local quasiparticle injection. 
We find that quasiparticle poisoning is dominated by the propagation of pair-breaking phonons across the chip. We characterize the energy dependence of the timescale for quasiparticle poisoning. Finally, we observe that incorporation of extensive normal metal quasiparticle traps leads to a more than order of magnitude reduction in quasiparticle loss for a given injected quasiparticle power.
\end{abstract}

\maketitle

Gate and measurement fidelities of superconducting qubits have reached the threshold for fault-tolerant operations 
\cite{Fowler12, Barends14}; however, continued progress in the field will require improvements in coherence and the development of scalable approaches to multiqubit control. Recently it was shown that nonequilibium quasiparticles (QPs) represent a dominant source of qubit decoherence \cite{Martinis09, Catelani11}. Quasiparticles are also a source of decoherence in topologically protected Majorana qubits \cite{Cheng12}. Most commonly, superconducting quantum circuits are operated in such a way that there is no explicit dissipation of power on the quantum chip; nevertheless, stray infrared light from higher temperature stages leads to a dilute background of nonequilibrium QPs in the superconducting thin films. According to \cite{Riwar16}, the leading mechanism for QP relaxation at low density $x = n_{\rm QP}/n_{\rm CP} \lsim x_*\simeq 10^{-4}$ is trapping by localized defects or vortices, where $n_{\rm QP}$ is the QP density and $n_{\rm CP}$ is the density of Cooper pairs ($4 \times 10^6$ $\mu$m$^{-3}$ in aluminum). In this regime, QPs propagate diffusively through the superconductor until they are trapped.

For future multiqubit processors, however, it might be necessary to integrate proximal classical control or measurement elements tightly with the quantum circuit, leading to a nonnegligible level of local power dissipation. For example, one approach to scalable qubit control involves manipulation of  qubits by quantized voltage pulses derived from the classical Single Flux Quantum (SFQ) digital logic family~\cite{McDermott14, Liebermann16}; here, local generation of QPs during each voltage pulse is inevitable.  Due to the local nature of dissipation, the QP density may become large, $x\gtrsim x_*$, and QP recombination accompanied by phonon emission to the substrate emerges as the leading mechanism of QP relaxation.  The emitted phonons can travel great distances through the substrate until they are absorbed by the superconducting film, leading to the generation of new QP pairs in remote regions of the film \cite{Eisenmenger67, Otelaja13}.

In this Letter, we present experiments to characterize the dynamics of QP poisoning in superconducting thin films subjected to direct QP injection, so that recombination is important and a significant flux of pair-breaking phonons is emitted to the substrate. 
We show that cuts in the superconducting film, which eliminate direct diffusion of QPs, have little influence on QP poisoning far from the injection site; however, the incorporation of normal metal QP traps leads to a suppression of QP poisoning by more than an order of magnitude. The data are well explained by a model where injected QPs recombine, emitting high-energy phonons that break pairs in distant parts of the chip. We study the energy dependence of the QP poisoning time and find it is consistent with the phonon-mediated mechanism. There have been prior attempts to suppress QP poisoning using trapped magnetic flux vortices \cite{Nsanzineza14, Wang14}; however, it can be challenging to trap a large number of vortices controllably while avoiding the microwave loss contributed by vortices themselves \cite{Song09}. Recently it was shown that incorporation of normal metal traps that are tunnel-coupled to the superconductor can enhance the QP removal rate by approximately a~factor of 4~\cite{Riwar16}.

\begin{figure}[t]
\includegraphics[width=.48\textwidth]{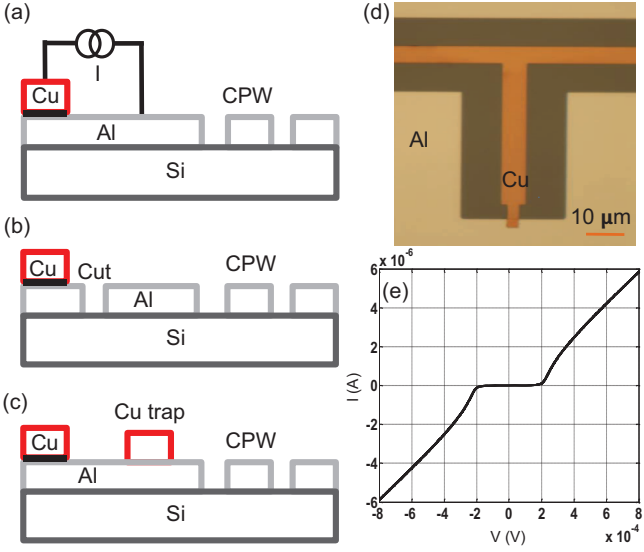}
\vspace*{-0.0in} \caption{Cross-sectional view of CPW resonators with NIS injectors for the three geometries studied: (a) direct injection into the groundplane, (b) groundplane cuts around the NIS injector, and (c) coverage of groundplane with normal metal QP traps. (d) Micrograph of the 10 $\mu$m$^2$ NIS (Cu/AlO$_x$/Al) QP injector. (e) Typical \textit{I-V} curve of the NIS injector.}
\label{fig:figure1}\end{figure}

\begin{figure}[t]
\includegraphics[width=.48\textwidth]{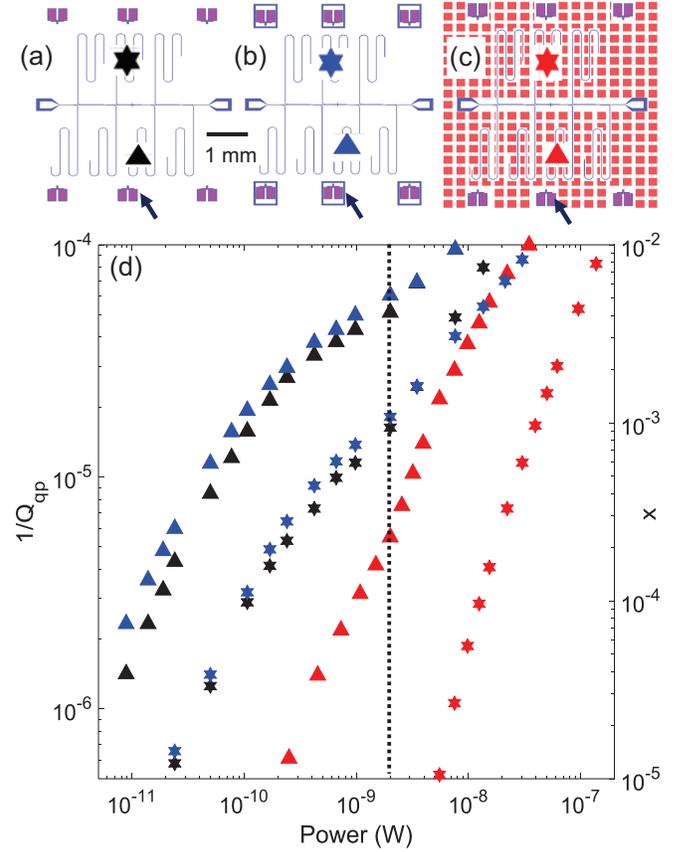}
\vspace*{-0.0in} \caption{Layout of multiplexed $\lambda/4$ CPW resonators with NIS injectors for the three geometries studied: (a) direct injection into the groundplane, (b) groundplane cuts around the NIS injector, and (c) coverage of groundplane with normal metal QP traps.  The NIS injectors used in these experiments are indicated by arrows. (d) Quasiparticle loss $1/Q_{\rm qp}$ and QP fraction $x$ \textit{versus} injected QP power for the three geometries studied (\textit{a}-black symbols, \textit{b}-blue symbols, and \textit{c}-red symbols). For each geometry, we plot the loss measured in nearby (triangles) and distant (stars) resonators. The bias voltage $3\Delta/e$ is indicated by the vertical dashed line.}
\label{fig:figure2}\end{figure}

In the experiments, we probe QP-induced loss in Al thin-film coplanar waveguide (CPW) resonator chips subjected to direct QP injection via normal metal-insulator-superconductor (NIS) tunnel junctions. Schematic cross-sectional views of the devices are shown in Fig.~\ref{fig:figure1}a-c. Each die consists of \rfm{seven} $\lambda/4$ CPW resonators capacitively coupled to a common feedline; \rfm{six} NIS junctions are arrayed around the chip perimeter. The device geometry enables investigation of the spatial variation of the instantaneous QP density for a given injection power. The resonators were fabricated from 100~nm-thick Al films grown by sputter deposition on $0.375\,\text{mm}$-thick oxidized Si wafers; the devices were patterned photolithographically and defined with a Transene wet etch. The Cu-AlO$_x$-Al NIS junctions were next formed in a liftoff process. An~ion mill was used to remove the native oxide of the Al prior to controlled thermal oxidation of the tunnel barrier, and the Cu counterlectrode was deposited by electron beam evaporation. The~junction areas were 10\,$\mu$m$^2$ with specific resistances of order $1\,\text{k}\Omega\cdot\mu\text{m}^2$. In Fig.~\ref{fig:figure1}d we show a micrograph of \rfm{an injector} junction, and in Fig.~\ref{fig:figure1}e we show a typical junction $I-V$ curve. We have investigated three geometries:

\begin{itemize}

\item{\textit{Geometry a} (Figs.~\ref{fig:figure1}a,  \ref{fig:figure2}a) - Here, QPs are injected directly into the groundplane of the resonator chip, and no mitigation steps are taken to limit QP diffusion or to trap QPs.}

\item{\textit{Geometry b} (Figs.~\ref{fig:figure1}b,  \ref{fig:figure2}b) - In these devices, the superconducting film into which QPs are injected is isolated galvanically from the groundplane of the resonators by 50\,$\mu$m-wide cuts. These cuts prevent the direct diffusion of injected QPs to the measurement region.}

\item{\textit{Geometry c} (Figs.~\ref{fig:figure1}c,  \ref{fig:figure2}c) - These devices incorporate a grid of $200\times200\,\mu\text{m}^2$ normal metal QP traps arrayed throughout the chip groundplane with an areal fill factor of 0.44. The traps are deposited as the last step of device fabrication; an~\textit{in situ} ion mill clean of the Al underlayer is performed prior to deposition of the 100~nm-thick Cu trap layer to ensure good metal-to-metal contact between the layers. The QP traps are set back by a standoff distance of approximately 50\,$\mu$m from any resonator features to ensure negligible microwave loss from proximity-induced suppression of the gap in the superconducting groundplane.}

\end{itemize}

Devices are cooled to 100~mK in an adiabatic demagnetization refrigerator and transmission across the resonators is monitored using standard homodyne techniques; devices are packaged using nonmagnetic connectors and screws to inhibit the nucleation of flux vortices in the superconducting films and a single layer of cryogenic mu-metal shields external magnetic fields. We fit the frequency-dependent transmission across the resonator in the quadrature plane and extract the internal and coupling-limited quality factors of the resonator as a~function of QP injection rate. We subtract the baseline internal loss (of order $10^{-6}$) measured in the absence of explicit QP injection from the total internal loss in order to determine QP loss~$1/Q_{\rm qp}$, which is proportional to QP density \cite{Martinis09}. In all cases, the microwave drive power is reduced to the point where the measured QP loss~$1/Q_{\rm qp}$ shows negligible sensitivity to small changes in microwave drive power~\cite{DeVisser14}; this power level corresponds to an equilibrium photon occupation in the resonators around $5\times10^4$.

For each device geometry, we characterize QP loss for two resonators: one resonator close to the injection point ($\sim\,$100$\,\mu\text{m}$ at nearest approach) and a second resonator far from the injector ($\sim\,$3$\,\text{mm}$ away); the locations of these resonators are marked with triangles and stars, respectively, in Fig. \ref{fig:figure2}. 
We plot QP loss~$1/Q_{\rm qp}$ and reduced QP density $x$ \textit{versus} injected QP power in Fig.~\ref{fig:figure2}d. The injector resistances are reasonably well matched (140, 140, and 156 $\Omega$ for geometries \textit{a}, \textit{b}, and \textit{c}, respectively), so that a given injected power corresponds to a nearly identical range of injection energies for all three samples.

For geometry~\textit{a} (black symbols), we observe the onset of significant dissipation in the nearby resonator as soon as the NIS injector is biased above the gap edge.
For the distant resonator, the onset of QP loss is much more gradual, reflecting a reduction in the efficiency of poisoning for the more distant resonators. 
For geometry~\textit{b} (blue symbols), there is no direct path for QPs to diffuse from the injection point to the resonators due to the presence of the groundplane cuts. Nevertheless, we observe levels of QP poisoning that are nearly identical to those seen in geometry~\textit{a}. The measured dissipation is clearly dominated by a~mechanism other than direct diffusion of QPs. For bias points close to the gap edge, QPs near the injection point recombine via emission of 2$\Delta$ phonons; these recombination phonons are capable of propagating through the dielectric substrate and breaking pairs at distant parts of the circuit, leading to excess microwave dissipation. The density of QPs in the injection region can be roughly estimated as $x_{\rm inj} \simeq I_{\rm inj}/(eDdn_{\rm CP})$, where  $I_{\rm inj}$ is the injection current, $D$~is the QP diffusion constant, and $d$~is the thickness of the superconducting film. For typical currents $I_{\rm inj} = 1$~$\mu$A just above the gap edge and diffusivity $D=20$~cm$^2$/s, we find $x_{\rm inj} \approx 8\times10^{-3} \gg x_* \simeq 10^{-4}$, so that recombination dominates over QP trapping at the injection site. The range of injected powers considered here is relevant to operation of an SFQ pulse generator, where a single SFQ junction undergoing phase slips at a rate of 5 GHz will dissipate approximately 1~nW. The facts that poisoning via phonon emission is dominant even at the lowest injection energies and that naive attempts to suppress poisoning by limiting diffusion are not effective are the first key conclusions of this work.

At higher biases, the injected QPs quickly relax via fast scattering processes to the gap edge, emitting athermal phonons. For bias voltages in the range $\Delta/e < V < 3\Delta/e$, these phonons do not have enough energy to break Cooper pairs; as a result, the fraction of injected power that is converted to pair-breaking phonons decreases as injection energy is increased beyond the gap edge.  For bias voltage $V>3\Delta/e$, however, relaxation of injected QPs to the gap edge is accompanied by emission of phonons with a broad range of energies extending above~2$\Delta$; a fraction of these phonons are capable of breaking pairs in remote regions of the chip. Indeed, we observe a clear enhancement in the QP loss as the injector bias is increased beyond $3\Delta/e$ (indicated by the vertical dashed line in Fig. \ref{fig:figure2}d). 

\begin{figure}[t]
\includegraphics[width=.48\textwidth]{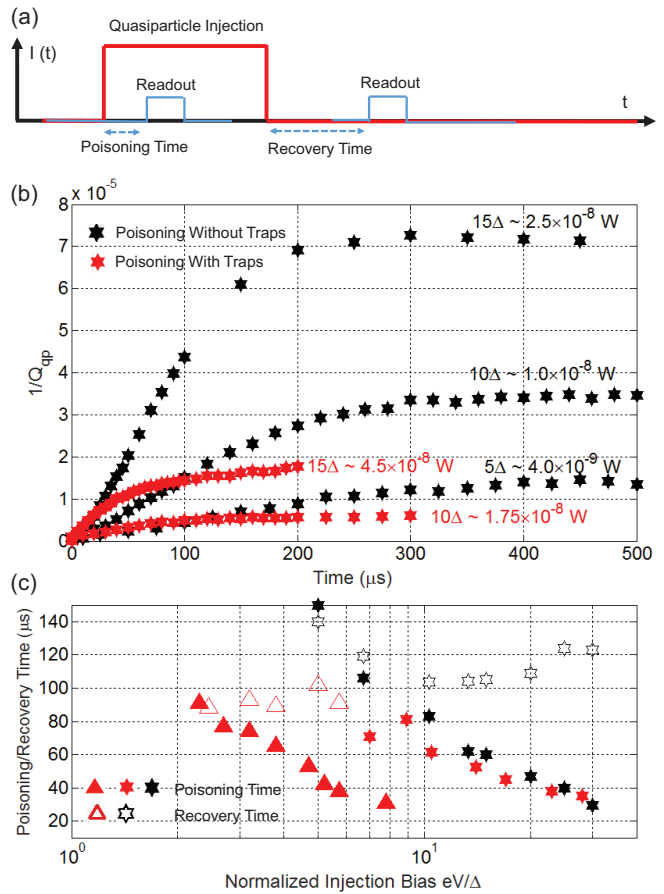}
\vspace*{-0.0in} \caption{(a) Experimental pulse sequence for the extraction of QP poisoning and recovery times $\tau_p$ and $\tau_r$, respectively. (b) Quasiparticle loss $1/Q_{\rm qp}$ \textit{versus} delay time following turn-on of QP injection pulse for devices without (black symbols) and with (red symbols) QP traps. Each data point is the result of a full frequency sweep through the resonance followed by correction for background dissipation. (c) Quasiparticle poisoning and recovery times $\tau_p$ and $\tau_r$ \textit{versus} normalized injection bias~$eV/\Delta$. Triangles correspond to resonators near the injection point, and stars correspond to far resonators (see Fig.~\ref{fig:figure2}a).}
\label{fig:figure3}\end{figure}


In the case of geometry~\textit{c} (red symbols), we find a more than order-of-magnitude suppression of QP loss for a given injected QP power for both the proximal and the distant resonators. It is expected that QPs that diffuse from the superconducting Al film to the normal metal traps will quickly lose most of their energy to conduction electrons in the normal metal via inelastic scattering \cite{Ullom2000}. 
Once QPs relax in the normal metal below the gap edge, they do not have enough energy to reenter the superconductor and hence are trapped. 
As phonon-mediated poisoning proceeds via multiple scattering events, each accompanied by the generation and recombination/relaxation of QPs, extensive coverage of the circuit groundplane with normal metal will limit the flux of pair-breaking phonons from the injection point to the measurement point. The effectiveness with which extensive normal metal coverage suppresses phonon-mediated QP poisoning is the second key conclusion of this work. 

Diffusion- and phonon-mediated poisoning should be readily differentiated by their dynamics, and we perform additional time-domain experiments to probe the characteristic timescales for QP poisoning and recovery. The experimental pulse sequence is shown in Fig.~\ref{fig:figure3}a and the time-dependent QP loss is shown in Fig.~\ref{fig:figure3}b.  The data reveals that dissipation in the resonator grows monotonically in time, reaching a~limiting value that depends on the injected power in a~manner that is consistent with our steady-state measurements. For each device, we fit the data with an exponential function and extract an energy-dependent QP poisoning time $\tau_{p}$. We find poisoning times of order 100~$\mu$s for the lowest injection energies, more than an order of magnitude shorter than the characteristic time for QP diffusion from the injection point to the measurement region. In a related experiment, we monitor the QP loss of the resonators as a~function of time following turn-off of QP injection, and we extract the characteristic QP recovery time $\tau_r$. In Fig. \ref{fig:figure3}c we plot poisoning and recovery times versus the normalized injection bias~$eV/\Delta$ for geometries \textit{a} and \textit{c} (with and without traps, respectively). In both cases, we find an~ approximate $1/V$ dependence of the QP poisoning time on the bias voltage. In contrast, there is no apparent voltage dependence for the recovery timescales. We understand that poisoning depends on the dynamics of phonon generation and propagation via multiple scattering events to the measurement region, and a~minimal model of phonon-mediated poisoning qualitatively captures the observed energy dependence of the poisoning time. By contrast, recovery is likely dominated by trapping and diffusion of low-energy QPs out of the center conductor of the $\lambda/4$ resonators, for which we expect little or no dependence on the injection energy.

We have developed a~simple model to describe the phonon-mediated QP poisoning of superconducting thin films. The model neglects the spatial dependence of the QP density, but treats the NIS injector and the measurement region separately. We follow the approach of Refs.~\cite{Martinis09, Lenander11, Wenner}, which is based on the Boltzmann kinetic equation with energy-dependent QP recombination and scattering rates \cite{Kaplan76}. Quasiparticle injection in the NIS region is described by a term that is proportional to the measured NIS current. We assume that injected QPs are uniformly distributed in a small volume around the NIS junction that is defined by the QP diffusion length or cuts in the superconducting film. We integrate the kinetic equation to determine the occupation of QP states and calculate the distribution of emitted phonons due to QP recombination and inelastic scattering.

Next, we use a modified Rothwarf-Taylor equation to analyze the reduced QP density $x$ at the measurement region \cite{Rothwarf67,Chang77}:
\begin{align}
r x^2+sx = g.
\end{align}
Here, $r$ and $s$ are the QP recombination and scattering rates, respectively, and $g$ is the rate of QP generation due to the flux of pair-breaking phonons from the injection site, calculated from the phonon distribution in the NIS region and scaled by a geometry-dependent factor characterizing the efficiency with which phonons propagate from the injection point to the measurement region. Note that in this naive model, the effect of adding normal metal traps is the same as increasing the distance from the injection point to the measurement point –- namely, to reduce the influx of pair-breaking phonons to the measurement region, as the traps strongly suppress poisoning via multiple pair-breaking and recombination/scattering events. A more refined model will take into account the energy dependence of phonon absorption and subsequent reemission between the injection and measurement regions.

\begin{figure}
  \includegraphics[width=0.49\textwidth]{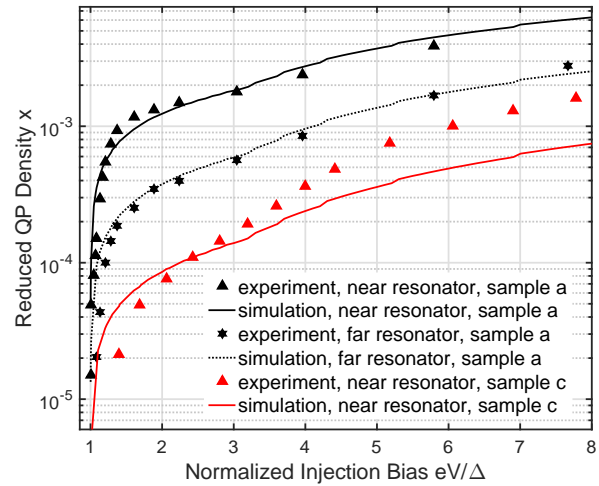}
    \caption{Steady-state QP density as a~function of normalized injection bias~$eV/\Delta$. Triangles and stars are the experimental values extracted from the measured quality factors $Q_{\rm qp}$~\cite{Martinis09} and lines are the simulated steady-state QP densities. The solid black line and black triangles correspond to the near resonator on sample \textit{a} (without normal metal traps), the dashed black line and stars correspond to the far resonator on the same device, and the red symbols and line correspond to the near resonator on sample \textit{c} (with normal metal traps). Simulation parameters are $s\simeq 7.7\times10^4$~s$^{-1}$ and $r\simeq 1.2\times 10^7$~s$^{-1}$. 
}
    \label{fig:figure4}

\end{figure}

In Fig.~\ref{fig:figure4}, we compare the simulation results to the steady-state QP densities extracted from experiment for three resonators: the near and far resonators with geometry~\textit{a} (without normal metal traps) and the near resonator with geometry~\textit{c} (with normal metal traps).  While the match to experiment is not perfect, the numerical model does capture the trends well enough to support the picture of phonon-mediated QP poisoning. The discrepancy between the simulated and measured QP density in the case of geometry \textit{c} suggests the importance of energy-dependent processes for phonon-mediated poisoning in the presence of normal traps; this is a subject of ongoing research. From the fit, we obtain $s\simeq 7.7\times10^4$~s$^{-1}$ and $r\simeq 1.2\times 10^7$~s$^{-1}$. In comparison with the values reported in Ref.~\cite{Wang14} ($r\simeq 10^7$~s$^{-1}$ and $s\simeq 7\times10^2$~s$^{-1}$), we find a similar recombination rate, but a much higher background trapping rate.  The QP generation rates used for these simulations are in the ratio $g_{a,n}:g_{a,f}:g_{c,n}=1:0.25:0.054$, where $g_{a,n}$ and $g_{a,f}$ refer to the near and far resonators for geometry \textit{a} and $g_{c,n}$ refers to the near resonator for geometry \textit{c}.


The approach of localizing the dynamics into two zero-dimensional regions is a~significant simplification, and a~more complete model would require proper introduction of a~spatial dependence to the QP density in a~way similar to Refs.~\cite{Wang14, Riwar16}. However, the key parameter that justifies our approach and differentiates our work from Refs.~\cite{Wang14, Riwar16} is the QP density $x_{\rm inj}$ at the injection point. While the QP injection rate in Ref.~\cite{Wang14} corresponds to an~effective current of $0.08\,\mu\text{A}$, the injection currents in our measurements span a range from 1-10~$\mu$A, corresponding to QP density at the injection site 1-2 orders of magnitude higher. Higher QP density at the injection site enhances QP recombination, which results in the appreciable emission of pair-breaking phonons. In contrast, the data of ~\cite{Wang14, Riwar16} appear to be in excellent agreement with a model where poisoning proceeds via QP diffusion.

In conclusion, we have performed a systematic study of dissipation due to nonequilibrium QPs in superconducting quantum circuits. We find that the dominant mechanism for QP poisoning is pair breaking mediated by high-energy phonons; moreover, this mechanism leads to strongly energy-dependent QP poisoning times. We further demonstrate that while diffusion-limiting cuts in the superconducting groundplane are not effective, extensive coverage of the superconducting film with normal metal traps provides a more than order-of-magnitude suppression of QP loss. Future devices may employ strategies to inhibit propagation of pair-breaking phonons to further reduce QP poisoning. For example, engineered discontinuities in the acoustic impedance at the superconductor-substrate interface could inhibit transmission of phonons into and out of the dielectric substrate \cite{Kaplan79}, thereby confining them to a small region of the superconducting film that is remote from sensitive quantum devices. These experiments suggest that superconducting quantum circuits can be made robust to modest levels of dissipation on chip, as might be required for the integration of large-scale quantum circuits with proximal classical control and measurement hardware.

\medskip

\begin{acknowledgments}
We acknowledge stimulating discussions with G. Catelani and L. I. Glazman. This work was supported by the U.S. Government under Grant W911NF-15-1-0248.
\end{acknowledgments}

\end{document}